# Multiple Breakup of Higher-Order Spatial Solitons

**Ori Katz[+], Yoav Lahini[+], and Yaron Silberberg***

*Department of Physics of Complex Systems, Weizmann Institute of Science, Rehovot 76100, Israel*

[*]*Corresponding author: yaron.silberberg@weizmann.ac.il*

[+]*These authors contributed equally to this work.*

We study the breakup of high-order spatial solitons propagating in an AlGaAs slab waveguide. We experimentally observe the breakup of such beams into multiple fragments and identify the mechanism of this breakup as the combined effect of two- and three-photon-absorption. We show that the multiple breakup persists even when the value of two-photon-absorption is reduced by an order of magnitude, owing to the high value of three-photon-absorption of AlGaAs at the half band-gap. The experimental results extend known mechanism of soliton breakup induced by two-photon-absorption, and agree well with numerical beam-propagation simulations.

Spatial solitons are self trapped, non diffracting light beams in which the nonlinear index profile balances diffraction. The first-order (N=1) soliton has a hyperbolic-secant beam profile which remains constant during propagation, and is referred to as the 'fundamental soliton' [1]. Higher order (N>1) solitons are bound solutions to the nonlinear Schrödinger equation that have energies which are $N^2$ times higher than that of the fundamental soliton. These bound higher-order solutions exhibit periodic evolution of the beam profile during propagation and are sometimes referred to as 'breathing-solitons' [2-3]. Whereas fundamental solitons are extremely stable to perturbations, it is well known that higher-order spatial solitons are not as robust and generally break up into two diverging fundamental solitons under the influence of perturbations such as linear or two-photon absorptions (2PA) [3-6]. In a complete analogy, higher-order *temporal* solitons, which are short pulses having high peak-powers, can break-up in the presence of absorption or due to numerous additional non-dissipative mechanisms such as high-order dispersion, Raman scattering and self-steepening [2, 7]. Such breakups can play an essential role in the process of supercontinuum generation in photonic crystal fibers, and in soliton fiber lasers [7-8].

In this letter we report experimental investigation of the breakup of higher-order (N>3) spatial optical solitons in a planar AlGaAs waveguide (Fig. 1). The planar waveguide limits the diffraction to one transverse direction, which allows effectively two-dimensional propagation and (1+1)-dimension soliton generation [6,9-10]. Our experimental setup is presented schematically in Fig. 1: A Gaussian light beam is injected into an 8mm long AlGaAs slab waveguide, using a 40X microscope objective. The AlGaAs slab waveguide has a core-height of $d=1.5\mu m$, with core/clad refractive index of $n_0$=3.3426 and $n_{clad}$=3.3123, respectively. A TE or TM mode excitation is chosen by means of a rotating half-wave plate, and the beam diameter in the x-axis is set using a cylindrical lens. The light source is an OPO (Spectra-Physics OPAL) pumped by a mode-locked Ti:Sapphire laser, producing 130fs pulses with 80MHz repetition rate, at a wavelength of 1530nm. The output beam-profile and average power are measured at the output facet of the slab by an infrared camera (Hamamatsu C5840) and a digital power meter. An additional CCD camera images the top facet of the AlGaAs slab, where multi-photon fluorescence enables direct monitoring of the propagation of high intensities beams [11].

In previous studies of soliton propagation in planar waveguides, a splitting into *two* diverging beams was observed as the beam power was increased above the fundamental soliton power [5-6]. This was intuitively explained by the action of 2PA at the point where the breathing N=2 soliton compresses: the nonlinear absorption reduces the power at the beam crest and causes its low intensity wings to split off and form two diverging fundamental solitons [6]. In our experimental setup, selecting a wide 52μm diameter input beam enables higher soliton-order (N>4) to be obtained with the experimentally available power, and for more complex dynamics to be observed. The input beam soliton order is given by: $N = \sqrt{\dfrac{P_{peak} n_2 a_0 k^2}{2 n_0 d}}$, where $P_{peak}$ is the input beam peak-power, $a_0$ is a measure of the beam width (=FWHM/1.76 for a sech profile), k is the wave vector in the medium and $n_2 = 1.6 \cdot 10^{-13}$ cm$^2$/W is the nonlinear refractive index [5,10].

A summary of the experimentally measured beam profiles at the slab output for a TE-polarized input beam at different average powers is presented in Fig. 3a. At low powers, when nonlinear self focusing effects are negligible, the Gaussian beam diffracts in a linear fashion. As the input power is increased, self focusing sets in and the spatial width of the output beam decreases until the output mode has collapsed to dimensions smaller than those of the input beam. As the beam power is further increased, the beam profile breaks into a two-peaked distribution, in agreement with previous observations for the N=2 soliton [5-6]. However, when the beam power is further increased, a distinct *three-peaked* beam profile emerges, which sustains its shape throughout the maximum experimentally available power. At the highest powers, we have managed to map the propagation of the beam inside the slab by directly observing the fluorescence pattern at the top of the slab waveguide. These results are presented in Fig 3a, showing an initial self focusing dynamics of the beam which results in high power densities, followed by a symmetric breakup into three fragments.

To further investigate the dynamics of the beam propagation leading to the observed three-fold splitting in the output beam profile, we have numerically simulated the beam propagation, taking into account diffraction and nonlinear self-focusing and absorptions mechanisms. Under these conditions the propagating electric field spatial envelope can be described by the nonlinear Schrödinger equation (NLSE) [6,9]:

$$\frac{\partial A}{\partial z} = i\left(\frac{1}{2k_0 n_0}\frac{\partial^2}{\partial x^2} + n_2 k_0 |A|^2\right) A - \frac{1}{2}\left(\alpha_1 + \alpha_2 |A|^2 + \alpha_3 |A|^4\right) A \quad (1)$$

where A(x,z) is the propagating electric field envelope, normalized such that $I(x,z)=|A(x,z)|^2$ is the peak intensity of the beam, and $k_0$ is the wave vector in vacuum. The intensity dependent absorption is given by the sum of linear and nonlinear absorptions (3PA) [9]: $\alpha(I) = \alpha_1 + \alpha_2 I + \alpha_3 I^2$. In AlGaAs: $\alpha_1$=0.1 cm$^{-1}$, $\alpha_3$=0.04 cm$^3$/GW$^2$, and $\alpha_2$=0.03 cm/GW for the TM mode and 0.3 cm/GW for the TE mode [10]. The effect of pulse temporal broadening due to group-velocity-dispersion (GVD) was considered numerically by an additional reduction of the pulse peak-power according to: $P_{peak}(z) = \dfrac{P_{peak}}{1 + z/(\tau_0^2/2\beta_2)}$, where $\tau_0$ is the initial pulse length, and $\beta_2$=1.35ps$^2$/m is the GVD coefficient [2,10]. The numerical simulation results were obtained by solving the NLSE using the split-step Fourier method [2], and were time-averaged to take into account the evolutions of the different temporal parts of the pulse, having different intensities [6]. Numerical results for the experimental parameters given in [10] are presented in Figs. 3b and 4b. It can be seen that the numerical results are in good agreement with the experimental results (Figs. 3a and 4a.) without the use of any fit parameter.

Further investigation into the role of 2PA versus 3PA in the multiple soliton breakup has been made by rotating the input polarization from TE to TM. This reduces the value of $\alpha_2$ tenfold, while leaving the other propagation parameters practically constant [10]. Experimentally measured beam-profiles for the TM mode show similar patterns to the ones observed for the TE mode, in accordance with numerical simulation results (not shown). These results suggest that 3PA is a significant mechanism in higher order soliton breakup in AlGaAS waveguides, in accordance with the high value of 3PA for AlGaAs at the half band-gap. To verify this conclusion, detailed numerical investigation of the N=5 soliton breakup process with different values of linear and nonlinear absorption coefficients has been done. Fig. 4 gives the summary for the simulated average-power evolution of the beam for a N=5 soliton, showing the part of the different dissipating mechanisms taking place during propagation and breakup process. A breakup of the breathing solution is obtained only when nonlinear 2PA or 3PA effects are present. The N=5 soliton breaks up into three fragments even when 3PA is the only dissipating mechanism (Fig. 3c).

Although multiple-breakup is a complex process resulting form the interplay of nonlinear effects, diffraction and dispersion, one can still have a qualitative understanding for the breakup mechanism. As the breathing soliton focuses, the energy-loss due to the nonlinear absorption increases. As a result the remaining power is too low to support the bound multisoliton state, which breaks up. The final outcome of this complex process depends on the exact propagation parameters and measuring point, as can be observed in the measured and simulated beam propagations (Fig. 3).

In summary we have presented experimental study of the breakup of high-order spatial soliton beams into multiple fragments in an AlGaAs slab waveguide. The results extend known mechanism of soliton breakup by 2PA, and agree well with numerical beam-propagation simulations.

**References**

1. G.I.A. Stegeman, D.N. Christodoulides, M. Segev, "Optical spatial solitons: historical perspectives", IEEE JSTQE 6, 1419-1427 (2000).

2. G.P. Agrawal, "Nonlinear Fiber Optics", Academic press.

3. V.V. Afanasjev, J.S. Aitchison, Y.S. Kivshar, "Splitting of high-order spatial solitons under the action of two-photon absorption", Opt. Comm. 116, 331-338 (1995).

4. Jaroslaw E. Prilepsky, B.I. Verkin, "Breakup of a multisoliton state of the linearly damped nonlinear Schrödinger equation", PRE 75, 036616 (2007).

5. Y. Silberberg, "Solitons and two-photon absorption," Opt. Lett. 15, 1005- (1990).

6. J. S. Aitchison, Y. Silberberg, A. M. Weiner, D. E. Leaird, M. K. Oliver, J. L. Jackel, E. M. Vogel, and P. W. E. Smith, "Spatial optical solitons in planar glass waveguides," J. Opt. Soc. Am. B 8, 1290- (1991).

7. M. G. Banaee and Jeff F. Young, "High-order soliton breakup and soliton self-frequency shifts in a microstructured optical fiber", J.Opt. Soc. Am. B/Vol. 23, No.7, 1484-1489 (2006).

8. D. Y. Tang, L. M. Zhao, B. Zhao, and A. Q. Liu, "Mechanism of multisoliton formation and soliton energy quantization in passively mode-locked fiber lasers", Phys. Rev. A 72, 043816 (2005); O. Katz, Y. Sintov, Y. Nafcha and Y. Glick, "Passively mode-locked Ytterbium fiber laser utilizing chirped-fiber-Bragg-gratings for dispersion control". Opt. Comm. 269, 156–165 (2007).

9. J. S. Aitchison, D. C. Hutchings, J. U. Kang, G. I. Stegeman, and A. Villeneuve, "The Nonlinear Optical Properties of AlGaAs at the Half Band Gap", IEEE J. Q. Elect., 33, 341-348 (1997).

10. Jin U Kang, George I Stegeman, Alain Villeneuve and J Stewart Aitchison, "AlGaAs below half bandgap: a laboratory for spatial soliton physics", Pure Appl. Opt. 5 583-594 (1996).

11. D. Mandelik, H. Eisenberg, Y. Silberberg, R. Morandotti and J.S. Aitchison, "Observation of mutually-trapped multi-band optical breathers in waveguide arrays", Phys. Rev. Lett., **90**, 253902 (2003).

Figure 1: Schematic layout of the experimental setup. The InGaAs slab waveguide limits the diffraction to one transverse direction (x-axis).

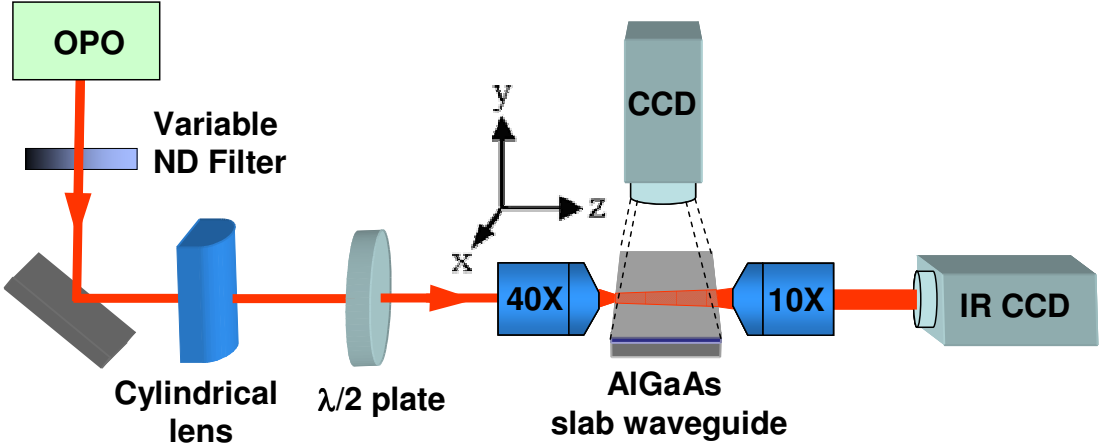

Figure 2: Summary of the beam profiles at the slab output for a TE-polarized input beam of 52μm diameter, at different input powers: (a) Experimentally measured profiles, y-axis depicts measured power at the slab output; (b) Numerically simulated profiles for the same input beam with the material parameters taken from [10].

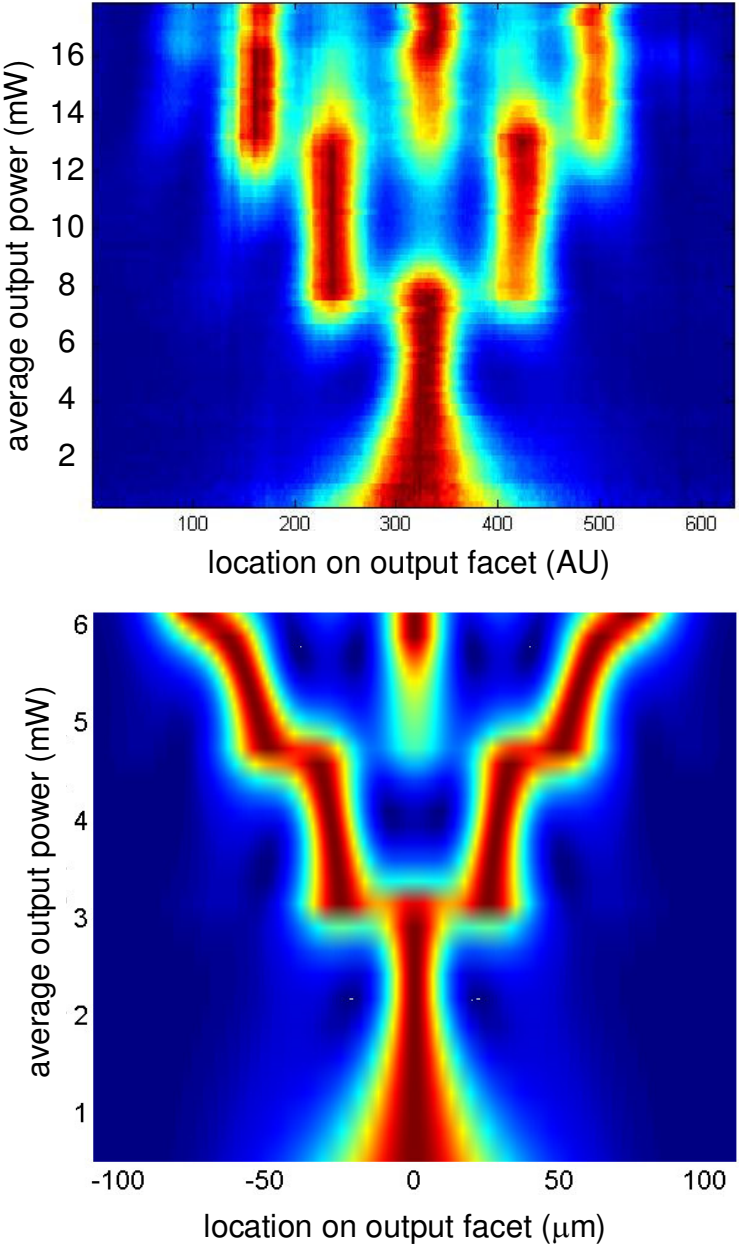

Figure 3: (a) Experimentally obtained image of the top facet showing fluorescence from propagation of a N~5 order soliton in a 8mm long AlGaAs slab (TE polarization); (b) Simulated propagation for N=5 soliton in a slab with the parameters taken from [10]; (c) Same as (b), but with 3PA as the only dissipation mechanism ($\alpha_1=\alpha_2=\beta_2=0$).

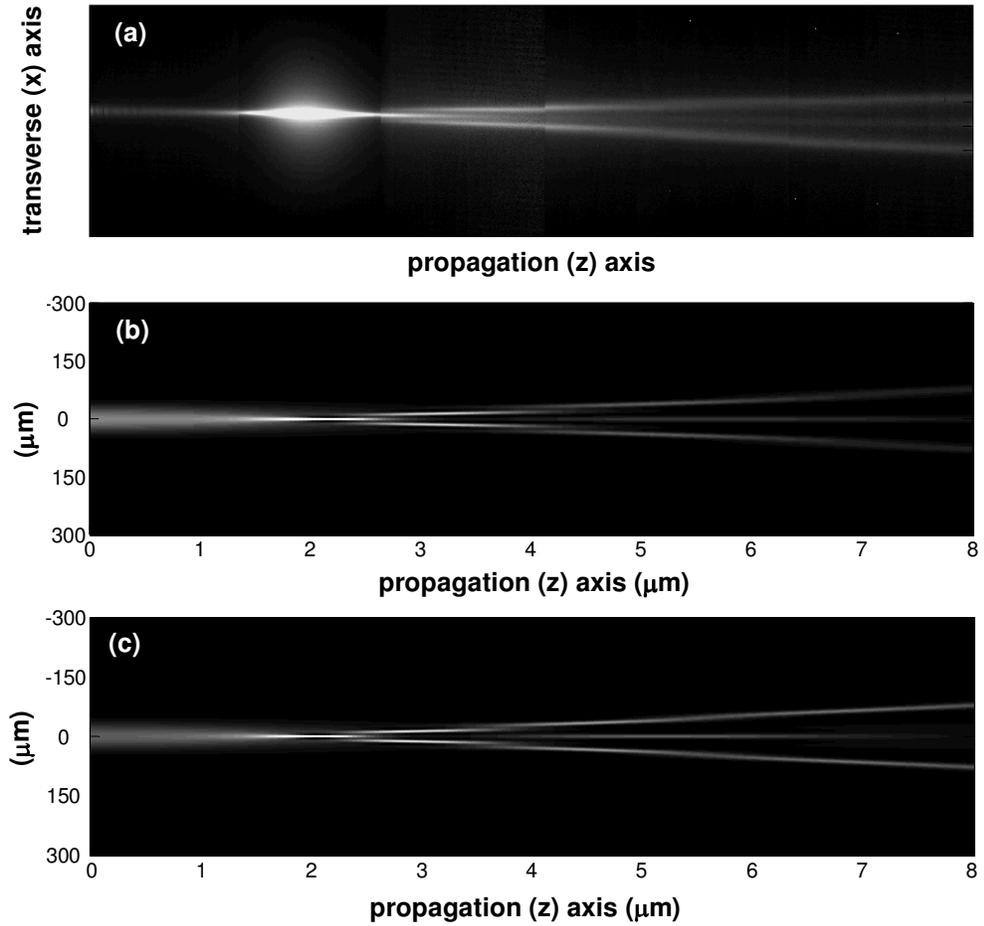

Figure 4: Numerically simulated evolution of the average beam power for an N=5 order soliton including different dissipating mechanism (the values of the different absorption coefficients are taken from [10] or are made equal to zero, accordingly).

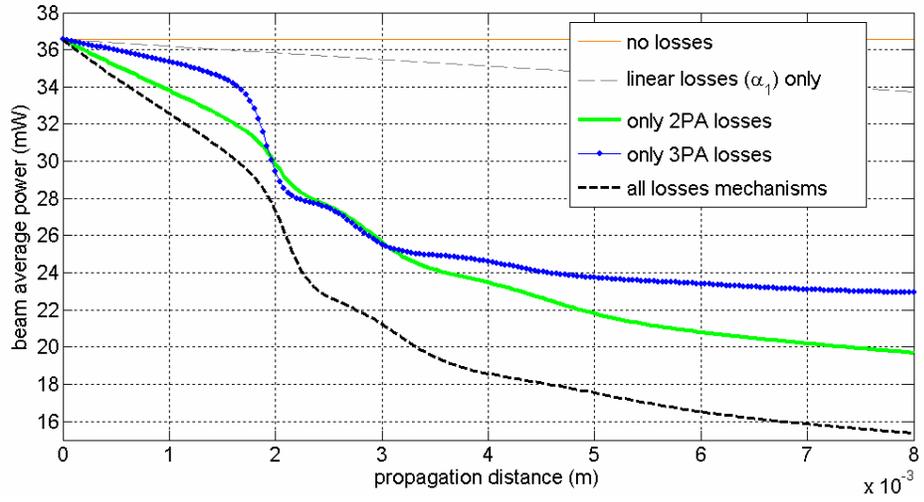